\documentclass[journal]{IEEEtran}

\usepackage{amsmath}
\usepackage{graphicx}
\usepackage{textcomp}
\usepackage{cite}
\usepackage{amssymb}
\usepackage{multirow}
\usepackage{url}
\usepackage{enumerate}
\usepackage{amsmath}  
\newtheorem{theorem}{Theorem}

\ifCLASSINFOpdf

\else

\fi

\newenvironment{shrinkeq}[1]
{ \bgroup
	\addtolength\abovedisplayshortskip{#1}
	\addtolength\abovedisplayskip{#1}
	\addtolength\belowdisplayshortskip{#1}
	\addtolength\belowdisplayskip{#1}}
{\egroup\ignorespacesafterend}

\begin{document}

\bstctlcite{IEEEexample:BSTcontrol}

\title{A Linear Solution Method of Generalized Robust Chance Constrained Real-time Dispatch}

\author{Anping Zhou,~\IEEEmembership{Student Member,~IEEE,}
	    Ming Yang,~\IEEEmembership{Member,~IEEE,}	    
	    and Zhaoyu Wang,~\IEEEmembership{Member,~IEEE}
	    
		\thanks{This work was supported by the National Science Foundation of China under Grant 51007047 and 51477091, and the Fundamental Research Funds of Shandong University.}		
		
		\thanks{A. Zhou and M. Yang (Corresponding Author) are with Key Laboratory of Power System Intelligent Dispatch and Control, Shandong University, Jinan, Shandong 250061, China (e-mail: myang@sdu.edu.cn).}
		\thanks{Z. Wang is with the Department of Electrical and Computer Engineering, Iowa State University, Ames, Iowa 50011, USA (e-mail: wzy@iastate.edu).}}
\maketitle

\begin{abstract}
In this letter, a novel solution method of generalized robust chance constrained real-time dispatch (GRCC-RTD) considering wind power uncertainty is proposed.
GRCC models are advantageous in dealing with distributional uncertainty, however, they are difficult to solve because of the complex ambiguity set. By constructing traceable counterparts of the robust chance constraints and using the reformulation linearization technique, the model is equivalently transformed into a deterministic linear programming problem, which can be solved efficiently by off-the-shelf solvers. Numerical results verify the effectiveness and efficiency of the approach.
\end{abstract}


\IEEEpeerreviewmaketitle

\section{Introduction}

\IEEEPARstart{T}{he} uncertainty of wind power introduces significant challenges to the real-time dispatch (RTD), which operates at a time-scale of minutes to determine the base points (BPs) and participation factors (PFs) of online units.
A variety of approaches, e.g., stochastic programming (SP) and robust optimization (RO), have been applied to address this problem. 
However, the effectiveness of the SP based approaches relies on the precise probability distribution of wind power, which is difficult to obtain in practice.
Meanwhile, the RO based approaches, which make decisions according to the bounds of disturbances, are usually criticized for their conservativeness.

The robust chance constrained dispatch approaches are proposed to fill the gap between the aforementioned two kinds of approaches.
A robust chance constrained optimal power flow (RCC-OPF) model and corresponding cutting-plane algorithm are proposed in \cite{lubin2016a}. In the model, the wind power forecast error (WPFE) is assumed to follow a normal distribution, and its first- and second-order moments are allowed to change within predetermined regions.
In \cite{bian2015distributionally}, a robust chance constrained model for reserve scheduling is developed, where the type of wind power distribution is not specified, but the moments are assumed to be known.
In \cite{Xie2016Distributionally}, the second-order cone programming is applied to solve the RCC-OPF model, where the expectation of WPFE must be 0 and the covariance matrix must be predetermined.
In practice, both the distribution type and moments are difficult to identify.
In \cite{zhang2017distributionally}, a generalized ambiguity set is used to capture uncertainties of renewable generations and load demands, which leads to a generalized robust chanced constrained (GRCC) OPF model. The model does not require a specific distribution type or precise moments, hence, it is more generic. However, the proposed semidefinite programming based algorithm is computationally intensive for online applications. 

The main contribution of this letter is to develop a fast solution method for the GRCC model so that it can be used for real-time dispatch, i.e., GRCC-RTD. The proposed method reduces the computational burden by constructing traceable counterparts of the robust chance constraints and applying the reformulation linearization technique (RLT).

\section{Problem Formulations}

Assume the mean vector and covariance matrix of WPFE vector $\boldsymbol{w}$ are $\boldsymbol{\mu}$ and $\boldsymbol{\Sigma}$, respectively,
and the statistical ones are $\boldsymbol{\mu}_0$ and $\boldsymbol{\Sigma}_0$.
Then, the model can be formulated as

\begin{shrinkeq}{-1.5ex}
\small{
\begin{flalign}\label{eq:1}
&Z=\min\limits_{\boldsymbol{p},\boldsymbol{\alpha}} \ {\boldsymbol{p}^\text{T}}{\boldsymbol{c}_1}\boldsymbol{p} + {\boldsymbol{c}_2}^{\text{T}}\boldsymbol{p} + {\boldsymbol{c}_3} \text{,}\\
&\text{s.t.} \ \ \ \ \ \ \, {\boldsymbol{e}^\text{T}}\boldsymbol{p} + {\boldsymbol{e}^\text{T}}{\boldsymbol{v}} - {\boldsymbol{e}^\text{T}}\boldsymbol{d} = 0 \text{,}  \\
&\ \ \ \ \ \ \ \ \ \  {\boldsymbol{e}^\text{T}}\boldsymbol{\alpha} = 1, \ \ 0\le \boldsymbol{\alpha} \le 1 \text{,}  \\
&\inf\limits_{\boldsymbol{w}\in D} \Pr\left( \underline{p}_i \le p_i - \alpha_i \boldsymbol{e}^\text{T}\boldsymbol{w} \le \overline{p}_i \right) \ge 1- \epsilon_{1,i}, \ \ \forall i \in G \text{,}\\
&\inf\limits_{\boldsymbol{w}\in D} \Pr\left( p_i^{d} \le \alpha_i \boldsymbol{e}^\text{T}\boldsymbol{w} \le p_i^{u} \right) \ge 1- \epsilon_{2,i} , \ \ \ \ \ \ \ \forall i \in G \text{,}\\
&\inf\limits_{\boldsymbol{w} \in D}\Pr \Big\{ \left| {{\boldsymbol{m}_{gl}^\text{T}}\left( {\boldsymbol{p} -  {\boldsymbol{e}^\text{T}}\boldsymbol{w}\boldsymbol{\alpha}} \right) + {\boldsymbol{m}_{wl}^\text{T}}\left( {{\boldsymbol{v}} + \boldsymbol{w}} \right) + {\boldsymbol{m}_{dl}^\text{T}}\boldsymbol{d}} \right| \le \overline{T}_l \Big\} \notag \\
& \ \ \ \ \ \ \ \ \ \ \ \ \ \ \ \ \ \ \ \ \ \ \ \ \ \ \ \ \ \ \ \ \ \ \ \ \ \ \ \ \ \ \ \,  \ge 1 - { \epsilon _l}, \ \ \forall l \in L \text{,} \\
&D=
\begin{cases}
{\int {f\left(\boldsymbol{w} \right)d\boldsymbol{w}}  = 1, \ \ f\left( \boldsymbol{w} \right) \ge 0} \text{,}\\
{{{\left[ {E\left( \boldsymbol{w} \right) - {\boldsymbol{\mu} _0}} \right]}^\text{T}}\boldsymbol{\Sigma} _0^{-1}\left[ {E\left( \boldsymbol{w} \right) - {\boldsymbol{\mu} _0}} \right] \le {\gamma _1}, \  {\gamma _1} \ge 0} \text{,}\\
{E\left[ {\left( {\boldsymbol{w} - {\boldsymbol{\mu} _0}} \right){{\left( {\boldsymbol{w} - {\boldsymbol{\mu} _0}} \right)}^\text{T}}} \right]}\preceq{{\gamma _2}{\boldsymbol{\Sigma} _0}, \ \ \ \ \ \ \, {\gamma _2} \ge 1} \text{,}
\end{cases}
\end{flalign}
}
\end{shrinkeq}

\noindent where $G$ is the set of online controllable units, e.g., units with automatic generation control;
$L$ is the set of transmission lines;
$D$ is the ambiguity set that determines the uncertainty level of WPFE;
$\boldsymbol{p}$ is the BP vector, and $p_i$ is the $i$th element of $\boldsymbol{p}$;
$\boldsymbol{\alpha}$ is the PF vector, and $\alpha_i$ is the $i$th element of $\boldsymbol{\alpha}$;
$\boldsymbol{c}_1$, $\boldsymbol{c}_2$ and $\boldsymbol{c}_3$ are the cost coefficient vectors;
$\boldsymbol{v}$ and $\boldsymbol{d}$ are the predicted wind power and load demand vectors;
$\overline{p}_i$ and $\underline{p}_i$ are the generation limits of unit $i$;
$p_i^u$ and $p_i^d$ are the adjustment limits of unit $i$;
$\epsilon_{1,i}$, $\epsilon_{2,i}$ and $\epsilon_{l}$ are the required risk levels;
$\gamma_1$ and $\gamma_2$ are the conservative coefficients;
$\boldsymbol{m}_{gl}$, $\boldsymbol{m}_{wl}$, and $\boldsymbol{m}_{dl}$ are the injection shift factor vectors;
$\overline{T}_{l}$ is the transmission limit of line $l$;
$\boldsymbol{e}$ is the vector of all ones;
and $f(\boldsymbol{w})$ is the joint probabilistic distribution function of $\boldsymbol{w}$.

The model in (1)-(7) is similar to the model in \cite{Xie2016Distributionally}. However, the ambiguity set in (7), which is adopted from \cite{zhang2017distributionally}, is more generic.
Besides, the constraints in (5) are added in the model to express the adjustment capability limits of the units.

\section{Solution Methodology}

In practice, the BPs and PFs should be updated very quickly.
However, the model in (1)-(7) is difficult to solve due to the existence of the robust chance constraints and the complexity of the ambiguity set.
To address this issue, the model has to be transformed.

Consider a robust chance constraint:

\setlength{\abovedisplayskip}{-4pt}
\setlength{\belowdisplayskip}{-4pt}
\begin{shrinkeq}{-0ex}
	\small{
		\begin{align}\label{eq:8}
            \mathop {\inf }\limits_{\boldsymbol{w} \in D} \Pr \left( {\boldsymbol{a}^\text{T}\boldsymbol{w} \le b} \right) \ge 1 - \epsilon \text{,}
		\end{align}
	}
\end{shrinkeq}

\noindent where $D$ is the set in (7).
Ref. \cite{zhang2016distributionally} provides a theorem to construct the deterministic counterpart of the constraint.

\setlength{\abovedisplayskip}{4pt}
\setlength{\belowdisplayskip}{4pt}
\begin{theorem}  
If $\gamma_1/\gamma_2 \le \epsilon$, (8) is equivalent to

\begin{shrinkeq}{-1ex}
	\small{
		\begin{align}\label{eq:9}
{\boldsymbol{\mu}_0 ^\text{T}}\boldsymbol{a}{+}\left( {\sqrt {{\gamma _1}}  + \sqrt {\left( {\frac{{1 -  \epsilon }}{ \epsilon }} \right)\left( {{\gamma _2} - {\gamma _1}} \right)} } \right)\sqrt {{\boldsymbol{a}^\text{T}}\boldsymbol{\Sigma}_0 \boldsymbol{a}}  \le b \text{;}
		\end{align}
	}
\end{shrinkeq}

Or else, (8) is equivalent to
\begin{shrinkeq}{-0ex}
	\small{
		\begin{align}\label{eq:10}
{\boldsymbol{\mu}_0 ^\text{T}}\boldsymbol{a}{+}\sqrt {\frac{{{\gamma _2}}}{ \epsilon }} \sqrt {{\boldsymbol{a}^\text{T}}\boldsymbol{\Sigma}_0 \boldsymbol{a}}  \le b \text{.}
		\end{align}
	}
\end{shrinkeq}  
\end{theorem} 

According to Theorem 1, the robust chance constraints in (4) and (5) can be directly transformed into equivalent deterministic linear constraints, regardless of the values of $\gamma_1$, $\gamma_2$, $\epsilon_1$, and $\epsilon_2$ (in fact, the constraints in (9) and (10) have the same structure). 
However, because the constraints in (6) are complex, their deterministic counterparts are quadratic.

For instance, assume all constraints in (4)-(6) satisfy the condition of (9).
They can be equivalently transformed into

\setlength{\abovedisplayskip}{-2pt}
\setlength{\belowdisplayskip}{-4pt}
\begin{shrinkeq}{-0ex}
	\small{
		\begin{align}\label{eq:11}
{\mu _s}{\alpha _i} + {k_{1,i}}{\alpha _i}\sqrt {{\Sigma _s}}  &\le {p_i} - {\underline{p}_i},{\kern 1pt} {\kern 1pt} {\kern 1pt} {\kern 1pt} {\kern 1pt} \forall i \in G \text{,}\\
- {\mu _s}{\alpha _i} + {k_{1,i}}{\alpha _i}\sqrt {{\Sigma _s}}  &\le {\overline{p}_i} - {p_i},{\kern 1pt} {\kern 1pt} {\kern 1pt} {\kern 1pt} {\kern 1pt} \forall i \in G \text{,}\\
{\mu _s}{\alpha _i} + {k_{2,i}}{\alpha _i}\sqrt {{\Sigma _s}}  &\le p_i^u,\ \ \ \ \ \ \ \forall i \in G \text{,}\\
- {\mu _s}{\alpha _i} + {k_{2,i}}{\alpha _i}\sqrt {{\Sigma _s}}  &\le  - p_i^d, \ \ \ \ \ \forall i \in G \text{,}
		\end{align}
	}
\end{shrinkeq}

\setlength{\abovedisplayskip}{-10pt}
\setlength{\belowdisplayskip}{-0pt}
\begin{shrinkeq}{-0ex}
\small{
\begin{align}\label{eq:15}
& k_l^2{\left( {{\boldsymbol{m}_{wl}} - \boldsymbol{e}\left( {\boldsymbol{m}_{gl}^\text{T}\boldsymbol{\alpha} } \right)} \right)^\text{T}}{\boldsymbol{\Sigma}_0}\left( {{\boldsymbol{m}_{wl}} - \boldsymbol{e}\left( {\boldsymbol{m}_{gl}^\text{T}\boldsymbol{\alpha} } \right)} \right) \le \notag \\ 
&{\left( {{T_{1,l}} - \boldsymbol{m}_{gl}^\text{T}\boldsymbol{p} + {\boldsymbol{\mu}_0}^\text{T}\left( {\boldsymbol{m}_{gl}^\text{T}\boldsymbol{\alpha} } \right)\boldsymbol{e}} \right)^2},\ \ \ \ \ \ \forall l \in L \text{,}\\
& k_l^2{\left( {{-\boldsymbol{m}_{wl}} + \boldsymbol{e}\left( {\boldsymbol{m}_{gl}^\text{T}\boldsymbol{\alpha} } \right)} \right)^\text{T}}{\boldsymbol{\Sigma}_0}\left( {{-\boldsymbol{m}_{wl}} + \boldsymbol{e}\left( {\boldsymbol{m}_{gl}^\text{T}\boldsymbol{\alpha} } \right)} \right) \le \notag \\ 
&{\left( {{T_{2,l}} + \boldsymbol{m}_{gl}^\text{T}\boldsymbol{p} - {\boldsymbol{\mu}_0}^\text{T}\left( {\boldsymbol{m}_{gl}^\text{T}\boldsymbol{\alpha} } \right)\boldsymbol{e}} \right)^2},\ \ \ \ \ \ \forall l \in L \text{,}
\end{align}
}
\end{shrinkeq}

\noindent where
$\mu_s = \boldsymbol{e}^\text{T}\boldsymbol{\mu}_0$;
${T_{1,l}} = {\overline{T}_l} - \boldsymbol{m}_{wl}^\text{T}\boldsymbol{v} - \boldsymbol{m}_{dl}^\text{T}\boldsymbol{d} - {\boldsymbol{\mu} _0}^\text{T}{\boldsymbol{m}_{wl}}$;
${T_{2,l}} = {\overline{T}_l} + \boldsymbol{m}_{wl}^\text{T}\boldsymbol{v} + \boldsymbol{m}_{dl}^\text{T}\boldsymbol{d} + {\boldsymbol{\mu} _0}^\text{T}{\boldsymbol{m}_{wl}}$;
$\Sigma_s = \boldsymbol{e}^\text{T}\left(\text{diag}(\boldsymbol{\Sigma}_0)\right)$;
${k_{1,i}} = \sqrt {{\gamma _1}}  + \sqrt {\left( {{(1 - { \epsilon _{1,i}})}/{{{ \epsilon _{1,i}}}}} \right)\left( {{\gamma _2} - {\gamma _1}} \right)}$;
${k_{2,i}} = \sqrt {{\gamma _1}}  + \sqrt {\left( {{(1 - { \epsilon _{2,i}})}/{{{ \epsilon _{2,i}}}}} \right)\left( {{\gamma _2} - {\gamma _1}} \right)}$;
and ${k_l} = \sqrt {{\gamma _1}}  + \sqrt {\left( {{(1 - { \epsilon _l})}/{{{ \epsilon _l}}}} \right)\left( {{\gamma _2} - {\gamma _1}} \right)}$.
 
\vskip 0.1cm
Therefore, the model in (1)-(7) can be equivalently transformed into a quadratically constrained quadratic programming (QCQP) problem, e.g., the model in (1)-(3) and (11)-(16).

To further simplify the model, the RLT is applied according to the structure of the transformed model.
Assume the decision vector $\boldsymbol{x} = [\boldsymbol{p}^\text{T},\boldsymbol{\alpha}^\text{T}]^\text{T}$, and let $\boldsymbol{X} = \boldsymbol{x}\boldsymbol{x}^\text{T}$.
Then, the transformed QCQP model can be rewritten as

\setlength{\abovedisplayskip}{1pt}
\setlength{\belowdisplayskip}{-2pt}
\begin{shrinkeq}{-1.5ex}
	\small{
		\begin{align}\label{eq:17}
		& Z = \min \  {\boldsymbol{Q}_0} \circ \boldsymbol{X} + \boldsymbol{b}_0^\text{T}\boldsymbol{x} + {c_0} \text{,}\\
		\text{s.t.} \ \ \ \ & {\boldsymbol{Q}_i} \circ \boldsymbol{X} + \boldsymbol{b}_i^\text{T}\boldsymbol{x} \le {c_i}, \ i \in I \text{,}\\
		&{\boldsymbol{Q}_j} \circ \boldsymbol{X} + \boldsymbol{b}_j^\text{T}\boldsymbol{x} = {c_j}, \ j \in M \text{,}\\
		& \boldsymbol{l} \le \boldsymbol{x} \le \boldsymbol{u} \text{,}
		\end{align}		
	}
\end{shrinkeq}

\noindent where $\boldsymbol{l}$ and $\boldsymbol{u}$ are the bounds of $\boldsymbol{x}$;
$I$ and $M$ are the inequality and equality constraint sets;
and $\boldsymbol{A} \circ \boldsymbol{B} = \sum\nolimits_{i,j = 1}^n {{A_{ij}}{B_{ij}}}$.

In the transformed QCQP model, the objective function and the deterministic counterparts of the transmission constraints are quadratic (see (1), (15) and (16)).
Thus, only $\boldsymbol{Q}_0$ of the objective function and $\boldsymbol{Q}_i$ of the transmission constraints are nonzero matrices, while $\boldsymbol{Q}_i$ and $\boldsymbol{Q}_j$ of other constraints are all zero matrices.
Moreover, it is also found that all the nonzero matrices can be expressed in symmetric forms.
For instance, $\boldsymbol{Q}_i$ corresponding to the transmission constraints in (15) can be expressed as

\setlength{\abovedisplayskip}{2pt}
\setlength{\belowdisplayskip}{2pt}
\begin{shrinkeq}{-1.5ex}
	\small{
		\begin{align}\label{eq:21}
    &{\boldsymbol{Q}_i} = \left[ {\begin{array}{*{20}{c}}
	\boldsymbol{A}&\boldsymbol{B}\\
	\boldsymbol{C}&\boldsymbol{D}
	\end{array}} \right] \text{,}
		\end{align}		
	}
\end{shrinkeq}

\noindent and the elements are

\begin{shrinkeq}{-1.5ex}
	\small{
		\begin{align}\label{eq:22}
		&{A_{ij}} = {A_{ji}} = \boldsymbol{m}_{gl}\left[ i \right]\boldsymbol{m}_{gl}\left[ j \right]\text{,}\\
		&{B_{ij}} = {C_{ji}} = {\boldsymbol{e}^\text{T}}{\boldsymbol{\mu} _0}\left( {\boldsymbol{m}_{gl}\left[ i \right]\boldsymbol{m}_{gl}\left[ j \right]} \right)\text{,}\\
		&{D_{ij}} = {D_{ji}} = \boldsymbol{m}_{gl}\left[ i \right]\boldsymbol{m}_{gl}\left[ j \right]\left[ {k_{l}^2\text{sum}\left({\boldsymbol{\Sigma} _0}\right) - {{\left( {\boldsymbol{e}^\text{T}}{\boldsymbol{\mu} _0} \right)}^2}} \right]\text{,}
		\end{align}		
	}
\end{shrinkeq}

\noindent where $\boldsymbol{A}$, $\boldsymbol{B}$, $\boldsymbol{C}$, $\boldsymbol{D}$ $\in \mathbb{R}^{n\times n}$; $n$ is the number of generators; $i$ and $j$ are indices from $1$ to $n$; $l \in L$; $\boldsymbol{m}_{gl}[i]$ represents the $i$th element of vector $\boldsymbol{m}_{gl}$; and $\text{sum}\left(\boldsymbol{\Sigma}_0\right)$ represents the sum of all elements in $\boldsymbol{\Sigma}_0$.

Therefore, the transformed QCQP model satisfies the precondition of applying the RLT, i.e., all matrices $\boldsymbol{Q}_0$, $\boldsymbol{Q}_i$ and $\boldsymbol{Q}_j$ are symmetric. 
According to RLT \cite{anstreicher2009semidefinite}, each element of $\boldsymbol{X}$, i.e., $X_{ij}$, can be treated as a new independent decision variable, and the QCQP problem can be transformed into a linear programming (LP) problem with the following auxiliary constraints:

\setlength{\abovedisplayskip}{2pt}
\setlength{\belowdisplayskip}{-6pt}
\begin{shrinkeq}{-1.5ex}
	\small{
		\begin{align}\label{eq:25}
		\boldsymbol{X} - \boldsymbol{l}{\boldsymbol{x}^\text{T}} - \boldsymbol{x}{\boldsymbol{l}^\text{T}} &\ge  - \boldsymbol{l}{\boldsymbol{l}^\text{T}} \text{,}\\
		\boldsymbol{X} - \boldsymbol{u}{\boldsymbol{x}^\text{T}} - \boldsymbol{x}{\boldsymbol{u}^\text{T}} &\ge  - \boldsymbol{u}{\boldsymbol{u}^\text{T}} \text{,} \\
		\boldsymbol{X} - \boldsymbol{l}{\boldsymbol{x}^\text{T}} - \boldsymbol{x}{\boldsymbol{u}^\text{T}} &\le  - \boldsymbol{l}{\boldsymbol{u}^\text{T}} \text{.}	
		\end{align}		
	}
\end{shrinkeq}

\section{Numerical Results}

The proposed solution method is tested on IEEE benchmark systems. The model is solved by MATLAB 2016a with CPLEX, on a PC with an Intel Core i5 CPU and 4 GB RAM. Unless otherwise specified, all risk levels, i.e., $\epsilon_{1,i}$, $\epsilon_{2,i}$ and $\epsilon_l$, are set to be 0.2, and the coefficients $\gamma_1$ and $\gamma_2$ are set to be 0.1 and 1.1, respectively.

A ``risk neutral'' model assuming there is no uncertainty and a Gaussian distribution based model assuming the distribution of WPFE is well known are adopted from [3] as benchmark models. The models are tested on the IEEE 118-bus system, where three wind farms are added at buses 17, 66 and 99, respectively. The maximum probability of constraint violations \cite{Xie2016Distributionally} according to the results of different models are summarized in Table \ref{Table_1}, where DRTD means the risk neutral model, GRTD means the Gaussian distribution based model, and GRCC means the GRCC-RTD model (in GRCC-1, $\gamma_1 = 0$, $\gamma_2 = 1$; in GRCC-2, $\gamma_1 = 0.1$, $\gamma_2 = 1.1$;  and in GRCC-3, $\gamma_1 = 0.2$, $\gamma_2 = 1.1$). WPFE samples generated from three different types of distributions, i.e., Gaussian distribution, Laplace distribution and logistic distribution, are used to perform the test.

\begin{table}[!htb]
    \newcommand{\tabincell}[2]{\begin{tabular}{@{}#1@{}}#2\end{tabular}}
	\renewcommand{\arraystretch}{1.2}
	\caption{Maximum Probability of Constraint Violations}
	\vspace*{-6pt}
	\label{Table_1}
	\centering
	\begin{tabular}{c| c c c c c}
		\hline
		\tabincell{c}{Distribution\\Type}& DRTD & GRTD & \tabincell{c}{GRCC-1} & \tabincell{c}{GRCC-2} & \tabincell{c}{GRCC-3} \\
		\hline
		Gaussian& 0.5031	& 0.2008	& 0.0209	& 0.0187	& 0.0171\\
		Laplace	& 0.5065	& 0.1903	& 0.0228	& 0.0211	& 0.0195\\
		Logistic& 0.5029	& 0.3205	& 0.0985	& 0.0832 	& 0.0789\\
		\hline
		Cost (pu)& 16.695	& 17.136	& 17.1918	& 17.2224 	& 17.2476\\
		\hline		
	\end{tabular}
\end{table}

From the test results, it is observed that the risk neutral model has the highest constraint violation risk, which is much higher than the required level (0.2 in the test). Meanwhile, GRTD assumes that the WPFE follows a Gaussian distribution. If the samples are generated from the assumed Gaussian distribution, the GRTD model can control the risk under the required level. If the samples are generated from other distributions, e.g., the logistic distribution, the risk may exceed the required level significantly, indicating that the chance constraints are invalid in this case.

When $\gamma_1 = 0$ and $\gamma_2 = 1$, the GRCC-RTD model becomes the same as the model proposed in \cite{Xie2016Distributionally}, in which the first- and second-order moments of WPFE are assumed to be known. The uncertainty level of the moments increases with the increase of  $\gamma_1$ and $\gamma_2$. From the results, it is observed that the higher the considered uncertainty level is, the lower the constraint violation risk will be, which indicates that GRCC-RTD can prepare appropriate reserve according to the moment uncertainty level to maintain the risk under the required level.  

It is seen from the table that the risk levels of the GRCC-RTD models are much lower than the required level for all three distribution types, which demonstrates the effectiveness of GRCC-RTD models in dealing with different uncertainty distributions. Meanwhile, all stochastic models, i.e., except the risk neutral one, have similar costs, which indicates that GRCC models can consider unspecific distribution types and imprecise moments without sacrificing the operational efficiency.

To illustrate the effectiveness of linearization, the costs and computation time of GRCC-2 with and without the RLT are listed in Table \ref{Table_2}. In the test, the QCQP model is also solved by CPLEX.

\vspace{-0.2cm}
\begin{table}[!htb]
	\newcommand{\tabincell}[2]{\begin{tabular}{@{}#1@{}}#2\end{tabular}}
	\renewcommand{\arraystretch}{1.2}
	\caption{Results with and without RLT}
	\vspace*{-6pt}
	\label{Table_2}
	\centering
	\begin{tabular}{c| c c}
		\hline
	   \ \ Model \ \	& cost (pu)	& Computation Time (s)\\
		\hline
		QCQP			& 17.2494	& 4.25\\
		LP				& 17.2224	& 2.31\\
		\hline		
	\end{tabular}
\end{table}

It is found that the cost of QCQP model is higher than that of the LP model, indicating the solution of QCQP may not be globally optimal. Meanwhile, it is found that 84\% more computation time is needed for solving the QCQP model.

\begin{figure}[!ht]
\setlength{\abovecaptionskip}{-0.1cm}
\setlength{\belowcaptionskip}{-0.cm}
	\centering
	\includegraphics[width=3.5in]{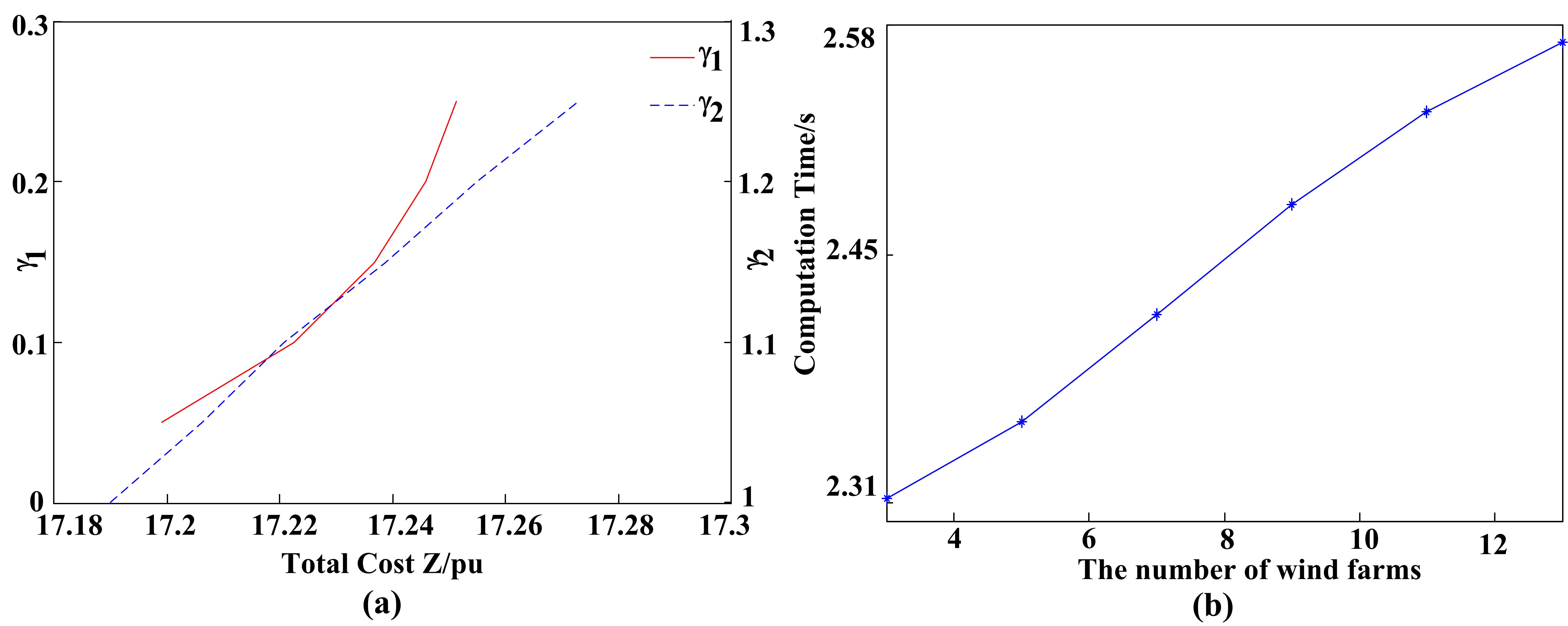}
	\caption{Sensitivity analyses in the IEEE 118-bus system.}
	\label{F1}
\end{figure}

To further test the proposed linear solution method, sensitivity analyses are performed on the 118-bus system, and the results are shown in Fig. \ref{F1}. 
Fig. \ref{F1}(a) illustrates the relationship between the conservative coefficients $\gamma_1$, $\gamma_2$ and the operational cost $Z$. It is observed that a higher $\gamma_1$ or $\gamma_2$ will lead to a higher $Z$. That is to say the more ambiguous the statistic result is, the more reserve should be prepared to maintain a low risk level, thus forcing the BPs moving away from the economic operating points and increasing the operational cost. However, the cost increase is not significant.

Fig. \ref{F1}(b) shows the computation time when different numbers of wind farms are connected to the system. As the number of wind farms increases from 3 to 15, the computation time slightly increases from 2.31s to 2.53s, which demonstrates the effectiveness of the proposed method in dealing with larger numbers of wind farms. Even for the case with 15 wind farms, the computation is still fast enough for online applications.

\begin{figure}[!ht]
	\setlength{\abovecaptionskip}{-0.1cm}
	\setlength{\belowcaptionskip}{-0.cm}
	\centering
	\includegraphics[width=2.3in]{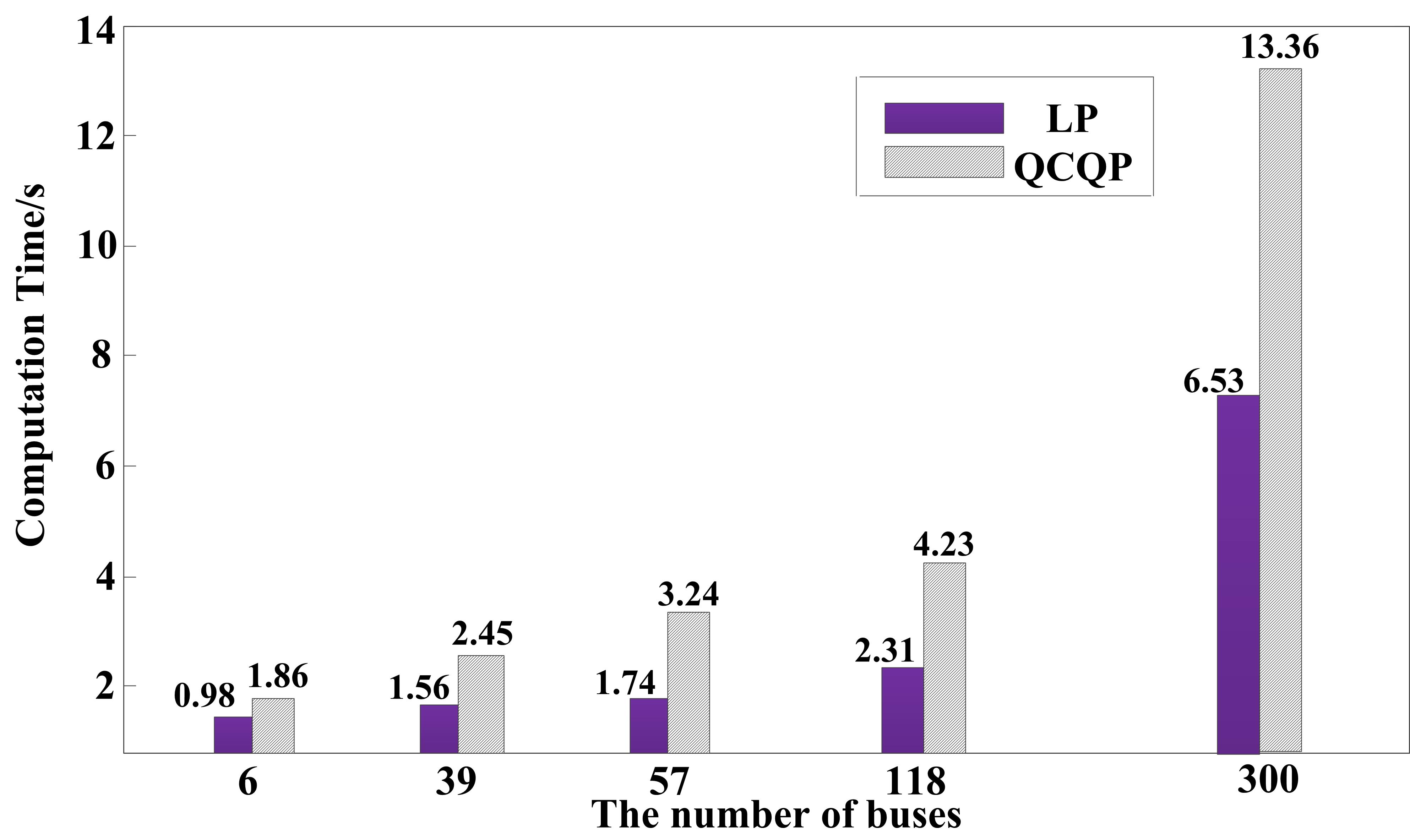}
	\caption{Computation time with different IEEE benchmark systems.}
	\label{F2}
\end{figure}

Fig. \ref{F2} summarizes the computation time on different IEEE benchmark systems, which further illustrate the efficiency of the proposed approach.

\section{Conclusions}

A novel linear solution method of GRCC-RTD is proposed in this letter.
The linearized model can maintain a low constraint violation risk while achieving relatively high operational efficiency.
Compared to the QCQP model, the linearized model can achieve global optimality with less computation time, revealing its potential application to large-scale power systems.


%

%
%
%
%
%

\ifCLASSOPTIONcaptionsoff
\newpage
\fi



\bibliographystyle{IEEEtran}
\bibliography{Reference}
\end{document}